\documentclass[aps,pre,twocolumn,floats]{revtex4}
\usepackage{epsfig}
\usepackage{bm}

\begin{document}

\newcommand{\hide}[1]{}
\newcommand{\tbox}[1]{\mbox{\tiny #1}}
\newcommand{\half}{\mbox{\small $\frac{1}{2}$}}
\newcommand{\sinc}{\mbox{sinc}}
\newcommand{\const}{\mbox{const}}
\newcommand{\trc}{\mbox{trace}}
\newcommand{\intt}{\int\!\!\!\!\int }
\newcommand{\ointt}{\int\!\!\!\!\int\!\!\!\!\!\circ\ }
\newcommand{\eexp}{\mbox{e}^}
\newcommand{\bra}{\left\langle}
\newcommand{\ket}{\right\rangle}
\newcommand{\EPS} {\mbox{\LARGE $\epsilon$}}
\newcommand{\ar}{\mathsf r}
\newcommand{\im}{\mbox{Im}}
\newcommand{\re}{\mbox{Re}}
\newcommand{\bmsf}[1]{\bm{\mathsf{#1}}}


\title{Quantum Dissipation due to the Interaction with Chaos}

\author{Doron Cohen$^{1}$ and  Tsampikos Kottos$^{2}$}

\affiliation{
$^{1}$ Department of Physics, Ben-Gurion University, Beer-Sheva 84105, Israel \\
$^{2}$Max-Planck-Institut f\"ur Str\"omungsforschung, Bunsenstra\ss e 10,
D-37073 G\"ottingen, Germany
}


\begin{abstract}
We discuss the possibility of having ``quantum dissipation" due to the interaction
with chaotic degrees of freedom. We define the conditions that should be satisfied
in order to have a dissipative effect similar to the one due to an interaction
with a (many body) bath. We also compare with the case where the environment
is modeled by a random matrix model. In case of interaction with ``chaos" we observe
a regime where the relaxation process is non-universal, and reflects the underlaying
semiclassical dynamics. As an example we consider a two level system (spin)
that interacts with a two dimensional anharmonic oscillator.
\end{abstract}

\maketitle


The interaction of a system with its environment is
a central theme in classical and quantum mechanics.
The main effects that are associated
with this interaction are ``dissipation" (irreversible loss of energy)
and ``noise". The latter is due to ``fluctuations"
of the environmental degrees of freedom.
On short time scales the main effect is
the ``decoherence" due to the noise.
On long time scales the interplay of dissipation and noise
leads to a state of thermal equilibrium.

The common modeling of the system-environment
interaction is provided by a Hamiltonian
${\cal H}_{total} = {\cal H}_0 + {\cal H}(Q,P; x)$,
where ${\cal H}_0$ is the system Hamiltonian,
$x$ is a system observable,
and ${\cal H}(Q,P; x)$ describes the
environment including the interaction
with the system. The simplest (and most
popular) modeling of the environment is
as a large collection of {\em harmonic oscillators}.
This is known as the Caldeira-Leggett approach \cite{weiss,leggett}.
Another approach is to use {\em random matrix} modeling
of the environment \cite{mello,pastur,esposito}.
However, in this paper we are interested
in another possibility, where the interaction
is with (few) {\em chaotic} degrees of freedom.

In what follows {\bf ``interaction with chaos"} means
that the environment is the quantized version
of a few degrees of freedom chaotic system.
This should be contrasted with {\bf ``interaction with bath"}
where the environment is modeled as a large
collection of quantized harmonic oscillators.
Let us regard the environment as a ``black box"
(one does not know what is there). The questions
that we would like to address are:
{\bf \ (1)} How to characterize the bath in a way
which does not assume a specific model.
{\bf \ (2)} Is it possible to distinguish ``interaction with chaos"
from ``interaction with bath".
As we explain below the second question
is related to the notions of ``thermodynamic limit"
and ``universality".

Common models for dissipation assume
a ``thermodynamic limit", which means
interaction with infinitely many degrees
of freedom. In this paper we inquire
whether {\em few} chaotic degrees
of freedom may have the same effect.
This is a question of great practical importance.
Future ``quantum electronics" may consist
of several interacting ``quantum dots".
One wonders how a coherent process
in one part of the ``circuit" is affected by the
"noise" which is induced by the quantized chaotic
motion of an electron in a nearby quantum dot.
In other words: one would like to know whether
it is possible to use in the nano-scale reality
notions such as ``dissipation" and ``dephasing",
that are traditionally associated with having an
interaction with many degrees of freedom.

As a specific example we consider a two level system (spin)
that interacts with a two-dimensional anharmonic oscillator.
This would be the well known ``Spin-Boson" model \cite{weiss,leggett}
if the interaction were with a bath of harmonic oscillators.
The motivation to deal with this model
is well documented in the cited literature.

We also compare with the case where the interaction
is with a {\em random-matrix modeled environment}.
In the ``quantum chaos"  literature, and in mesoscopic
physics, random matrix theory (RMT) is regarded
as the ``reference" case. Any deviation from RMT is called
"non-universality", and has to do with the underlying
semiclassical dynamics. In this paper we show
that the notion of (non) universality can be extended
into the studies of quantum dissipation.

The basic parameters that characterize {\em any}
system-environment modeling are listed in Table~1.
We shall define these parameters in a way which
is independent of the theoretical modeling of the bath.
A common assumption is going to be that the mean level spacing
is very small. For completeness of presentation,
and for the purpose of defining what does it mean ``very small",
we keep $\Delta$  as an explicit free parameter (note \cite{rmrk}).


\vspace*{0.2cm}

\begin{tabular}{|l|l|}
\hline
{\small \em parameter} & {\small \em \ \ \ \ significance} \\
\hline
$\Delta\propto\hbar^d$  &  environment mean level spacing \\
$\Delta_b\propto\hbar$ &  environment bandwidth  \\
$T$ &  environment temperature \\
$d_{\tbox{T}}$  &  environment heat capacity \\
\hline
$\EPS$  &  system energy \\
$A$ &  system amplitude of motion \\
$V$  &  system rate of motion \\
\hline
$\Gamma$  &  strength of coupling  \\
\hline
\end{tabular}

{\footnotesize \ \\ Table 1: The various parameters that
characterize a generic quantum dissipation problem.
(See text for details).}

\vspace*{0.0cm}
\pagebreak



As a leading example we consider a two level system.
For the system Hamiltonian we write
${\cal H}_0 =  (\hbar\Omega/2) \bm{\sigma}_1$
where $\Omega$ is the Bloch frequency,
and $\bm{\sigma}_{i=1,2,3}$ are the Pauli matrices.
One can think of this Hamiltonian  as
describing a particle in a double well potential \cite{weiss,leggett}.
Then it is natural to define its position as $x = v \bm{\sigma}_3$,
where $v$ is a constant. We assume that the interaction with the
environment is via this ``position" coordinate:
${\cal H}(Q,P;x) =
\half(P_1^2{+}P_2^2 + Q_1^2{+}Q_2^2) + (1{+}x) Q_1^2 Q_2^2$.
This environment can be interpreted as a particle moving
in a 2-dimensional anharmonic well (2DW).
In the representation $|\nu,n \rangle$,
which is determined by ${\cal H}(Q,P;0)$
and $\bm{\sigma}_3$, the Hamiltonian matrix takes the form
\begin{eqnarray} \label{e7}
H_{total} \ = \
\left[
\matrix{\bm{E}+v\bm{B} & \hbar\Omega/2 \cr
\hbar\Omega/2 & \bm{E}-v\bm{B}}
\right]
\end{eqnarray}
where $\bm{E}=\mbox{diag}\{E_n\}$ is a diagonal
matrix that contains the energy levels of the environment,
and $\bm{B}$ is a banded matrix (see \cite{lds}).
The initial state of the total Hamiltonian $\Psi(t=0)$
is assumed to be factorized as $\varphi \otimes \psi$,
where $\varphi$ is the initial state
of the spin, and $\psi=|n_0\rangle$
is the initial state of the environment.
It is implicit that we average over states
with $E_{n_0} \sim E$ corresponding to
a microcanonical preparation.
As for the spin, we would like to consider
the standard scenario where the initial
state is a coherent superposition
$|\varphi\rangle = ( |\uparrow\rangle + |\downarrow\rangle )/\sqrt{2}$.
The reduced probability matrix after time $t$ is
\begin{eqnarray} \label{e8}
\rho_{\nu,\nu'}(t)  =
\sum_{n} \Psi_{\nu,n}(t)^{*}\Psi_{\nu',n}(t)
\equiv \half (1+ \vec{M} {\cdot} \vec{\bm{\sigma}})_{\nu,\nu'}
\end{eqnarray}
where $\vec{M}=(M_1,M_2,M_3)$ is the
polarization of the spin. It is most convenient
to describe the state of the spin using $\vec{M}$.
In particular we define
$S(t) =\vec{M}\cdot\vec{M}$ as a
measure for the purity of the spin state.


We turn to formulate the general case.
The interaction of the system with the
environment is assumed to be of the general form
${\cal H}_{int}=-x{\cal F}$ where
$x$ and ${\cal F}$ are system and environmental
observables respectively. (In the above example
$x = v \bm{\sigma}_3$ and ${\cal F}=-Q_1^2 Q_2^2$).
In the absence of system-environment coupling
we can characterize the fluctuations of the
observable ${\cal F}(t)$ by a correlation
function $C(\tau)$.
(We are using here Heisenberg picture language).
Its Fourier transform $\tilde{C}_{\tbox{E}}(\omega)$ 
is known as the power spectrum of the fluctuations. 
The observable ${\cal F}$ has a matrix representation 
$\langle n | {\cal F} | m \rangle = -\bm{B}_{nm}$
where we use the basis which is determined by ${\cal H}$.
The fluctuations are related to the {\em bandprofile} 
of this matrix:
\begin{eqnarray} \label{e1}
\tilde{C}_{\tbox{E}}(\omega) &=& 
\left[\sum_{m}
\left|\bm{B}_{nm}\right|^2
\ 2\pi\delta\left(\omega-\frac{E_m{-}E_n}{\hbar}\right)\right]_{E_n {\sim} E} 
\\ \label{e2}
&\equiv& 2\pi\sigma^2\delta(\omega)+
\frac{2\pi\hbar\sigma^2}{\Delta} R\left(\frac{\hbar\omega}{\Delta}\right)
G\left(\frac{\hbar\omega}{\Delta_b}\right) 
\end{eqnarray}
In the first expression there is an implicit microcanonical averaging
over the states $E_n \sim E$. In the second expression $\sigma^2$ 
is the average value  $\overline{|\bm{B}_{nm}|^2}$,
taken over the near-diagonal matrix elements.
The lower cutoff function $R()$
depends on the level spacing statistics \cite{ophir}:
It is $R() \approx 1$ for $\omega > \Delta/\hbar$.
The mean level spacing $\Delta$ is proportional to
$\hbar^d$, where $d$ is the the number of environmental 
degrees of freedom.
Any {\em chaotic motion} is characterized by 
a finite correlation time $\tau_{c}$. Therefore
$\tilde{C}_{\tbox{E}}(\omega)$ has a a cutoff frequency 
$\omega_{c} = 2\pi/\tau_{c}$. This implies 
(via Eq.(\ref{e1})) that $\bm{B}_{nm}$ is a banded
matrix with a bandwidth $\Delta_b=\hbar\omega_c$.
The envelope function $G()$, with $G(0)\equiv1$,
describes the bandprofile.

For sake of comparison we refer to the {\em Spin-Boson}  
model. The distribution of the bath oscillators 
is described \cite{leggett,weiss} by a spectral function $J(\omega)$,  
leading to $\tilde{C}_{\tbox{E}}(\omega){=}2\hbar J(\omega)/(1{-}\eexp{-\beta\hbar\omega})$, 
where $\beta$ is the reciprocal temperature of the bath.
The ``ohmic" assumption of having ``white noise" 
($C(\tau)$ with short correlation time $\tau_c$)
for high temperatures is imposed {\em by construction}, 
by setting $J(\omega)\propto\omega G(\omega/\omega_c)$. 
This corresponds to a {\em strong chaos assumption}.
However, it should be realized that even if we ``cook" 
a harmonic bath that has the same $\tilde{C}_{\tbox{E}}(\omega)$ 
as that of a chaotic environment, still there is 
a major difference: In spite of having the same 
bandprofile, the $\bm{B}_{nm}$ matrix of a harmonic 
bath is very {\em sparse}: only states that differ 
by ``one photon" excitation of a {\em single} 
oscillator are coupled (hence the differences $E_m{-}E_n$
are the frequencies of the harmonic oscillators).

In what follows we would like to define the notion
of temperature ($T$) without assuming a specific modeling.
There are three features of the environment that has to do 
with this notion: {\bf (i)} The growing density of states; 
{\bf (ii)} The growing fluctuations intensity;   
{\bf (iii)} The asymmetry of $\tilde{C}_{\tbox{E}}(\omega)$ with 
respect to $\omega$. 
Let us regard the system as a ``thermometer".
The equilibrium is determined by the microcanonical 
temperature $T=(\partial_E \ln(1/\Delta))^{-1}$. 
For $d{=}2$ environment with constant density of states 
we would get $T{=}\infty$. [We consider this hypothetical 
case for argumentation purpose. This would be indeed 
the case if the environment were modeled as a billiard. 
In a later paragraph we discuss the model 
of Eq.(\ref{e7}) for which $T$ is finite].
In case of a two level system, having $T=\infty$   
implies an equal probability for the two energy states.  
Now we can take (instead of a two level system) 
a particle with one degree of freedom 
as a ``thermometer". Since we already know that the 
``temperature" is (say) $T=\infty$, we may deduce that 
the friction coefficient, as determined by 
the fluctuation-dissipation theorem is $\eta = \nu/(2T) = 0$, 
where $\nu=\tilde{C}_{\tbox{E}}(\omega{\sim}0)$ is the 
intensity of the fluctuations. 
This conclusion is {\em wrong}. 
In fact the friction coefficient
is $\eta = \half\Delta {\times} \partial_E(\nu/\Delta)$, 
as discussed in \cite{frc} (and references therein). 
Thus we can define an effective temperature, 
which is related to the thermalization process: 
\begin{eqnarray} \label{e3}
T_{\tbox{eff}}  \ \equiv \ \frac{\nu}{2\eta} \ = \ 
\left(\frac{\partial}{\partial E}
\ln\left(\frac{1}{\Delta} \tilde{C}_{\tbox{E}}(\omega{\sim}0) \right)
\right)^{-1}
\end{eqnarray}
In generic circumstances the distinction between 
$T_{\tbox{eff}}$ and $T$ is not so dramatic. 
In fact, some further inspection into Eq.(\ref{e1})
reveals that generically $T_{\tbox{eff}} = T/2$. 
The above subtlety does not apply to a thermal preparation. 
Assuming a reciprocal temperature~$\beta$, 
the friction coefficient (as obtained by canonical averaging 
over the above cited microcanonical result) is $\eta = \half \beta \nu$. 
Therefore we get $T_{\tbox{eff}}{=}1/\beta$ as expected.

Having defined $T$, we can make a quantum mechanical 
distinction between high and low temperature regimes.
This is related to the asymmetry of $\tilde{C}_{\tbox{E}}(\omega)$
with respect to $\omega$. 
Some inspection into Eq.(\ref{e1}) reveals that 
a finite temperature implies that the band profile acquires 
a factor $\exp(\omega/(2T_{\tbox{eff}}))$,
which is consistent with the Spin-Boson modeling 
(see a previous paragraph). 
{\em ``High temperature"} means that for
the physically relevant frequencies $\hbar\omega/T\ll 1$.
A~sufficient condition is $T\gg\Delta_b$.
In the latter case $\tilde{C}_{\tbox{E}}(\omega)$ can be treated as
a symmetric function with respect to $\omega\mapsto-\omega$,
and therefore it can be interpreted as the power
spectrum of a {\em classical noise}.

The issue of {\em thermodynamic limit} is related 
to the heat capacity of the environment. This is    
$d_{\tbox{T}} = ({\partial T}/{\partial E})^{-1} \sim d$,
where $d$ is its number of {\em environmental} degrees of freedom.
The energy that the system can exchange with the environment
is denoted by $\EPS$.
If we want to assume a stable temperature $T$,
the heat capacity of the environment should be large enough,
so that energy exchange between the system
and the environment does not have a big effect.
This leads to the condition $\EPS \ll  d_{\tbox{T}} \times T$.
For a generic few degrees of freedom system
$\EPS \sim  d_0 \times T$,
where $d_0$ is the number of {\em system}  degrees of  freedom.
This leads to $d_0 \ll  d_{\tbox{T}}$.
In case of a two level system the condition is much easier.
Namely, we have $\EPS \sim \hbar\Omega$, and therefore
we get the ``easy" condition $\hbar\Omega \ll  d_{\tbox{T}} \times T$.

Assuming the typical circumstances of
an oscillating system, the time variation of the system
observable $x(t)$ is characterized by an amplitude $|x| \sim A$,
and by a rate of change $|\dot{x}| \sim V$.
This specification of the system dynamics is essential
in order to define a dimensionless parameter
that characterizes the system-bath interaction:
\begin{eqnarray} \label{e5}
& \frac{\Gamma}{\Delta} \ \ = \ \ \mbox{minimum}\left(
\frac{2\pi\sigma^2}{\Delta^2}A^2, \ \left(\frac{\hbar\sigma}{\Delta^2}V\right)^{2/3}\right)
\end{eqnarray}
Loosely speaking this parameter indicates how many
environmental energy levels are mixed non-perturbatively
due to the interaction with the system. The $A$ dependence
of $\Gamma$ is the consequence of the well known theory by Wigner:
It is the number of levels which are mixed by the perturbation
in ${\cal H} \mapsto E_n\delta_{nm} + x \bm{B}_{nm}$ with $|x|\sim A$.
If the perturbation is slow (small $\dot{x}$) this $A$ based
estimate becomes non-relevant. For a proper analysis \cite{frc}
one should switch to the adiabatic ($x$ dependent) basis,
leading to ${\cal H} \mapsto E_n\delta_{nm} + \dot{x} (i\hbar\bm{B}_{nm}/(E_n{-}E_m))$
with $|\dot{x}|\sim V$. The $V$ based estimate for $\Gamma$
follows from the latter representation. Thus one realizes that
for slow rate of $x$ variation there is a crossover form
a $A$-determined to a $V$-determined mixing.


The rest of this paper is aimed in clarifying the significance
of the parameter $\Gamma$. Some results of the simulations
with the 2DW model Eq.(\ref{e7}) are presented in Fig.1.
The energy of the environment was in the range
$2.8<E<3.2$ where the classical dynamics is predominantly chaotic.
The classical correlation time is $\tau_{c}\sim1$.
The simulations are done with $\hbar=0.03$.
This means that the bandwidth is $\Delta_b \sim 0.2$.
This should be contrasted with the mean level spacing $\Delta\sim 0.004$.
The temperature in the specified energy range is $T\sim 1.3$,
and the heat capacity is $d_{\tbox{T}}\sim2.4$.
The amplitude of the motion is $A=v$,
while the rate of the motion is formally $V=\infty$.
The latter should be understood in 
the path-integral context (Eq.(\ref{e9})),
where $x(t)$ makes ``jumps" between the two sites ($x=\pm v$).
Hence $\Gamma\propto(v/\hbar)^2$.
One can easily verify that the Kondo parameter \cite{leggett}
is $\alpha = (1/16\pi)\Gamma/T$. In our simulations $\alpha \ll 1$.

In the lower panels of Fig.1 we present
the corresponding results of simulations
with a RMT model where the induced fluctuations
have {\em exactly} the same power spectrum
as in the 2DW model. The RMT model has been
obtained by taking the Hamiltonian Eq.(\ref{e7})
with a randomized $\bm{B}$.
We simply randomized the signs of the off-diagonal
elements. This procedure destroys all the
correlations between the elements, but does
not affect the bandprofile (which is implied by Eq.(\ref{e1})).
Two observations should be made
immediately: {\bf (i)} A few degrees of freedom chaos
indeed provides a dissipative effect,
as in the case of a many degrees of freedom bath.
{\bf (ii)} The effect of interaction with ``chaos"
can be distinguished from the case of
a random-matrix modeled environment.
The latter claim is based on the observation that
in the regime $v>0.16$ there is a two orders
of magnitude difference between the
corresponding curves. Also the scaling of curves
with respect to $v$ is ``broken" (upper panel):
The sensitivity to $v$ is much smaller than implied
by the overcompensating scaling of the time axis.


As explained in the introduction the deviation from RMT
is regarded as ``non-universality" and its understanding
requires a proper definition of the {\bf classical limit}.
This should not be mistaken as a synonym for
"high temperatures". The confusing, and possibly meaningless
procedure to define this limit is by taking $\hbar\rightarrow0$.
We argue below that {\em the meaningful definition
of a ``semiclassical regime" is by considering the
nature of the dynamics of the environment}.
This leads to the condition $\Gamma \gg \Delta_b$.
In order to explain  this condition we adopt
the Feynman-Vernon  picture of the dissipation process.
Within this picture the propagator of the
reduced probability matrix
is written as a path integral:
\begin{eqnarray} \label{e9}
K(\nu,\nu'|\nu_0,\nu_0') =
\sum_{x_A,x_B} F[x_A, x_B]
\ \eexp{i({\cal A}[x_A]-{\cal A}[x_B])}
\end{eqnarray}
The summation is over pairs of system
trajectories (with weight factors absorbed into the
definition of the integration measure).
The action ${\cal A}[x]$ is defined as the phase
which is accumulated along a given trajectory.
We note that for a spin the trajectory is piecewise
constant ($x=\pm v$).
The influence functional is defined as
%
$F[x_A, x_B] =
\langle \psi |
U[x_B]^{-1} U[x_A]
|\psi \rangle$
%
where the expectation value is taken for the
initial preparation of the environment
(it is typically a mixture of many states
implying that an appropriate average should
be taken over $\psi$).
The environmental evolution operator in case
of the models that we consider is
$U[x] = {\cal T}\exp
\left( -(i/\hbar)\int_0^t (\bm{E}+x(t')\bm{B})dt'\right)$.
Thus, in a semiclassical framework the problem
of ``quantum dissipation" reduces to that
of analyzing ``{\em driven degrees of freedom}".
It has been realized \cite{crs} that the 
driven dynamics becomes non-perturbative
if $\Gamma>\Delta_b$. Below we discuss the implication 
of this observation.

The purity $S(t)$ is related to the dephasing factor $|F[x_A, x_B]|$.
[For $\Omega{=}0$, the Hamiltonian Eq.(\ref{e7}) becomes block diagonal,
leading to $S(t) {=} |F[x_A, x_B]|^2$, with $x_A{=}v$ and $x_B{=}{-}v$].
If the fluctuations are regarded as ``noise",
one obtains the {\em standard} expression
%
$|F[x_A,x_B]| = \eexp{-\frac{1}{2\hbar^2}\int_0^t\int_0^t
C(t'{-}t'')(x_B(t')-x_A(t''))^2 dt'dt''}$.
%
This expression implies a short time 
Gaussian decay $S(t)=\exp(-4 C(0) (vt/\hbar)^2)$,   
which evolves into a long time Gaussian decay
$S(t)=\exp(-(\sigma v t / \hbar)^2)$
in the regime $\Gamma < \Delta$.  
If the envelope $G()$ of the bandprofile 
were smooth, then we would expect,   
in the regime $\Delta < \Gamma < \Delta_b$,
an intermediate stage of exponential decay 
$S(t)=\exp(-2\gamma t)$ with
$\gamma=2 (v/\hbar)^2 \tilde{C}_{\tbox{E}}(\Omega)\sim \Gamma/\hbar$. 
In case of our numerical example the bandprofile has a structure \cite{lds},
and hence $C(\tau)$ has oscillations that show up in the simulations.

Can we trust the standard expression for $|F[x_A,x_B]|$
if we have a dynamical environment 
rather than a noise source [with the same $C(\tau)$]?
It is not difficult to observe that the standard expression   
can be derived from Fermi-golden-rule (FGR) considerations.  
In case of Harmonic bath the FGR treatment is valid, 
and this expression, with $C(\tau)$ replaced by its
symmetrized version, is {\em exact}.
But for interaction with "chaos" we have claimed above 
that non-universality should show up 
in the non-perturbative regime ($\Gamma > \Delta_b$).
Our numerics confirm this prediction:  
In the non-perturbative regime ($\Gamma > \Delta_b$) 
we find a premature ($t < \tau_c$) crossover 
from the expected short time Gaussian decay,  
to a non-universal behavior which is not captured 
by the standard formula. In the 2DW case the system has a
classical limit, and therefore the nonperturbative decay of $S(t)$
is slowed down (compared with RMT) because it is limited by the classical dynamics.
As observed (note the inset) the decay becomes much less sensitive to $v$.

The above discussed non-universality can be regarded
as the manifestation of ``semiclassical" correlations
between the off-diagonal matrix elements of $\bm{B}_{nm}$.
These are {\em not} reflected in $\tilde{C}_{\tbox{E}}(\omega)$.
In case of the {\em Spin-Boson} model the $\bm{B}_{nm}$ matrix 
is sparse, which implies (in the limit of infinite bath) 
that off-diagonal correlations can be neglected.
In case of a {\em random matrix} modeled environment,
absence of correlations is guaranteed by construction 
for $x=0$,  which implies (assuming no sparsity) 
lack of ``invariance" with respect to $x$ \cite{kuzna}.
Accordingly we have three classes of models that
become distinct in the non-perturbative regime.


In conclusion, we have discussed the consequences
of having four energy scales $(\Delta, \Delta_b, T, \Gamma)$ 
in any generic problem of ``quantum dissipation". 
The strength of the interaction is characterized by $\Gamma$.
A universal "quantum dissipation" behavior
requires a separation of energy scales
$\Delta \ll \Gamma \ll \Delta_b$.
Non-perturbative breakdown of this universality
due to the underlaying semiclassical dynamics
is found if $\Gamma > \Delta_b$.

We thank M. Esposito, P. Gaspard, L. Pastur, and  V.~Falco for stimulating discussions.
This work was supported by a Grant from the GIF,
the German-Israeli Foundation for Scientific Research and Development,
and by the Israel Science Foundation (grant No.11/02).



\clearpage
\onecolumngrid

\ \\

\epsfig{figure=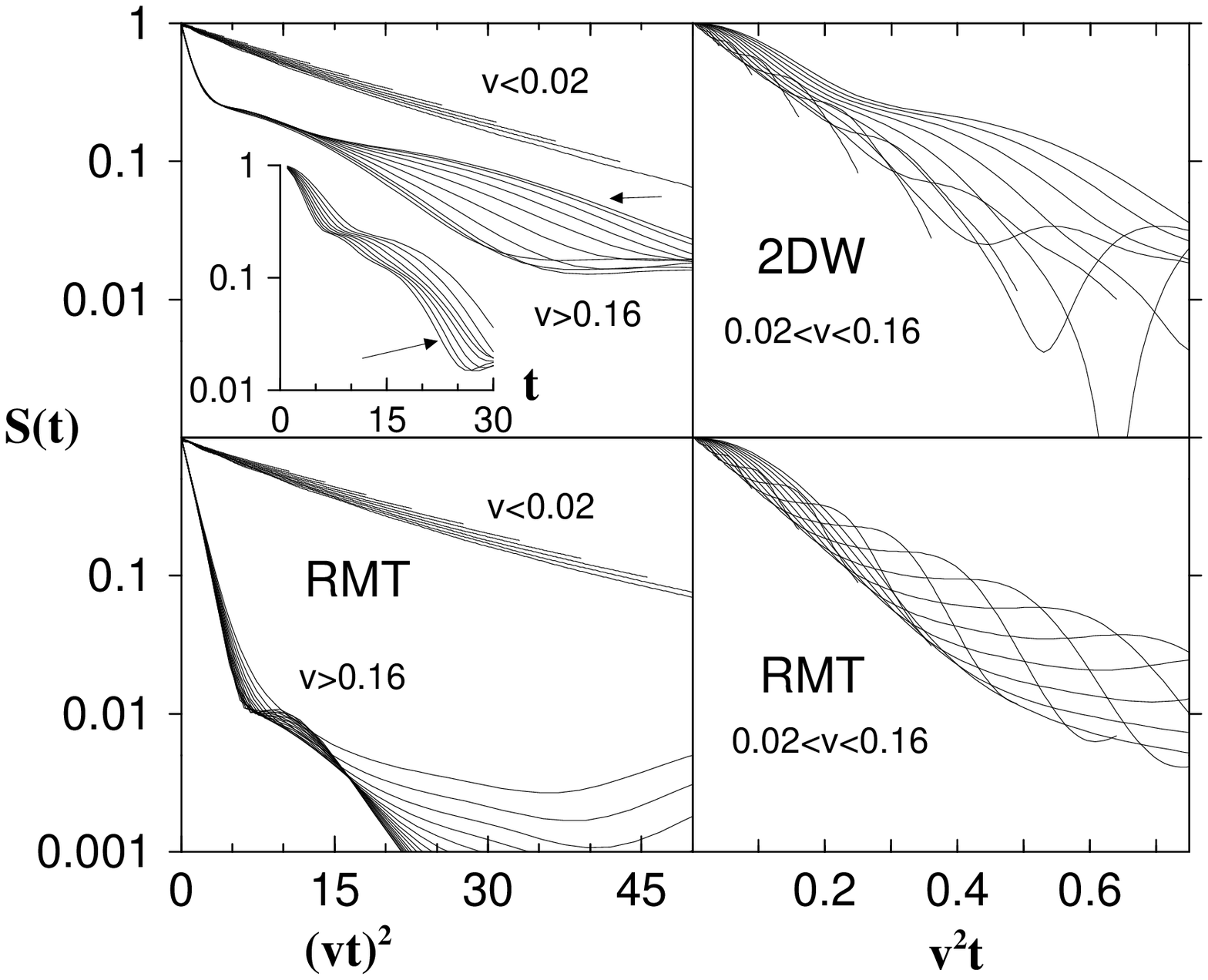, width=0.85\hsize}

\ \\

{\footnotesize\noindent {\bf FIG.1:} 
The decay of $S(t)$ in case of interaction with
chaos (upper panels) and in case of interaction with a ``randomized"
environment that has exactly the same fluctuations (lower panels).
The selected values of the coupling parameter are in the range \mbox{$10^{-4}<v<0.3$}.
The left panels are for the $\Gamma<\Delta$ regime \mbox{($v<0.02$)} and for the
$\Gamma>\Delta_b$ nonperturbative regime \mbox{($v>0.16$)}. 
The inset is for some of the \mbox{$v>0.16$} curves without
scaling. The arrows point on the largest $v$ value. 
The right panels are for the $\Delta<\Gamma<\Delta_b$ regime \mbox{($0.02<v<0.16$)}.
The scaling methods of the time axis are implied by the analysis of $S(t)$,
which should hold if there is a universal behavior which
is determined by $C(\tau)$ alone (see text).}


\begin{thebibliography}{99}

\bibitem{weiss}
{\em Quantum Dissipative Systems}, U. Weiss, World Scientific, Singapore (1999).

\bibitem{leggett}
A.J. Leggett et al, Rev. Mod. Phys. {\bf 59}, 1 (1987).

\bibitem{pastur} J. Lebowitz and L. Pastur,
preprint (2002).

\bibitem{esposito}
M. Esposito and P. Gaspard,
Phys. Rev. E {\bf 68}, 066113 (2003); 
Phys. Rev. E {\bf 68}, 066112 (2003).

\bibitem{mello}
P. Pereyra, J. Stat. Phys. {\bf 65}, 773 (1991).
P.A. Mello, P. Pereyra and N. Kumar, J. Stat. Phys. {\bf 51}, 77 (1988).

\bibitem{rmrk}
From here on we ignore the effect
of recurrences that happen after an extremely large time.

\bibitem{ophir}
O.M. Auslaender and S. Fishman,
J. Phys. A  {\bf 33}, 1957 (2000);
Phys. Rev. Lett. {\bf 84}, 1886 (2000).

\bibitem{frc} D. Cohen, Annals of Physics 283, 175 (2000).

\bibitem{lds}
D. Cohen and T. Kottos, Phys. Rev. E 63, 36203 (2001).

\bibitem{crs}
D. Cohen, Phys. Rev. Lett. {\bf 82}, 4951 (1999).
D. Cohen and T. Kottos, Phys. Rev. Lett. {\bf 85}, 4839 (2000).


\bibitem{kuzna}
Attempts to have an $x$ invariant random matrix model, [such as in
P. A. Bulgac et al, PRE {\bf 58}, 196 (1998)],
make the model ``perturbative by construction" \cite{crs}.

\end{thebibliography}
\end{document}